\documentstyle[preprint,eqsecnum,aps]{revtex}
\begin{document}
\draft
\preprint{HYUPT-94/13
         \hspace{-28.5mm}\raisebox{2.4ex} {SNUTP 94-129}}
\title{BRST Cohomology and Its Application to QED}
\author{Hyun Seok Yang and Bum-Hoon Lee}
\address{Department of Physics, Hanyang University, Seoul 133-791, Korea}
\maketitle

\begin{abstract}
\hspace{.5cm}
We construct the BRST cohomology under a positive-definite inner product
and obtain the Hodge decomposition theorem at a non-degenerate state vector
space $V$. The harmonic states isomorphic with a BRST
cohomology class correspond to the physical Hilbert space with positive
norm as long as the completeness of $Q_{BRST}$ is satisfied.
We explicitly define a ``co-BRST'' operator and analyze the quartet
mechanism in QED.
\end{abstract}


\narrowtext
\section{INTRODUCTION}
\label{sec:intro}

The covariant quantization of constrained systems and the renormalization of
gauge theories heavily depend
on the BRST approach \cite{Henn92,Baul85}. This approach
extends the phase space, including the anticommuting ghost
variables. Thus, we must project out all the physical states in a
positive-definite Hilbert space in order to recover the probabilistic
interpretation of the quantum theory.
We achieve this goal by asking for the BRST invariant
states of ghost number zero \cite{Naka90}.
However, the nilpotency of the BRST operator
gives the equivalence classes of physical states, which naturally leads us
into the cohomology group interpretation about the physical Hilbert space.
This use of the BRST cohomology to characterize the physical states can be
strengthened more by introducing the adjoint operation of
the BRST operator \cite{Nish84,Spie87,Kala91}.
This statement is analogous to the Hodge decomposition theorem of
differential geometry, which naturally leads to an isomorphism between the
space of the cohomology group and the space of
harmonic forms \cite{Eguchi}.
The physical Hilbert space can be identified with the harmonic
states which have positive-definite norms as long as the disastrous paired
singlet with non-zero ghost number is absent. Plausible
arguments exist for the absence of singlet pairs in the actual models of
gauge theories \cite{Naka90}. Following Refs. 4 and 5,
we will show the apparent correspondence between the irreducible
representation of the BRST algebra and the BRST cohomology.

In Sec. II, we construct the BRST cohomology under a positive-definite
inner product defined on a non-degenerate state space graded with
a ghost number and obtain the Hodge decomposition theorem.
In order to define a positive-definite inner
product, we introduce the metric on the state space $V$, which is the
analogue of the Euclidean complex conjugation {\bf C} in
Ref. 5 and the time reversal operation $T$ in Ref. 8.
This metric matrix provides us an isomorphism between the subspaces
$V_{p}$ and $V_{-p}$ of the respective ghost numbers $+p$ and $-p$,
and this isomorphism induces an isomorphism in the BRST cohomology.
The metric matrix $\eta$ plays an important role in the proof of the
quartet mechanism and the ``split a pair'' mechanism.
{}From the metric matrix, we easily find that $Tr\;\eta$ is equal to the
dimension of the BRST singlet subspace $V_S$ with a positive-definite norm.
The new positive definite inner product provides us the orthogonal
decomposition of the state space $V$. This decomposition into a sum of
linearly independent subspaces is just the Hodge decomposition theorem.
We will show how the illuminating example in Ref. 5 can be
realized through the quartet mechanism.

In Sec. III, we explicitly analyze the
BRST cohomology of QED using the mode expansion of the field operators.
We find the ``co-BRST'' operator defined in Sec. II and show the quartet
mechanism and the pair-splitting mechanism taking an analogy with Ref. 5.

\section{BRST COHOMOLOGY}
\label{sec:BRST}

\def\be{\begin{equation}}
\def\ee{\end{equation}}
\def\bea{\begin{eqnarray}}
\def\eea{\end{eqnarray}}
\def\ba{\begin{array}}
\def\ea{\end{array}}

Let us assume that a state vector space $V$ with an indefinite metric is
non-degenerate and equipped with complete basis vectors $\{|h_j>\}$.
Then, an arbitrary vector $|x>$ in $V$ can be represented uniquely as
\be
|x>=\sum_j x_j |h_j>.\label{stax}
\ee
The metric matrix $\eta$ on the vector space $V$ is defined by
\be
\eta_{ij}=<h_i|h_j>,\label{metric}
\ee
which is Hermitian,
\be
\eta^{\dagger}=\eta,\label{herm}
\ee
and the non-degeneracy of the space $V$ can be
expressed as
\be
det\; \eta\neq 0.\label{det}
\ee
The representation $t$ on $V$ of a linear operator $T$ is defined by
\be
T|h_j>=\sum|h_k>t_{kj}.\label{reT}
\ee
The matrix representations $t$ and ${\bar t}$
of $T$ and $T^{\dagger}$, respectively, should satisfy
the condition \cite{Nish84}
\be
\eta{\bar t}=t^{\dagger}\eta.\label{mrt1}
\ee
Here, $t^{\dagger}$ is the Hermitian conjugate of the matrix $t$.
In particular, when the operator $T$ is Hermitian, that is
$T^{\dagger}=T$, we have
\be
\eta t=t^{\dagger}\eta \label{mrt2}
\ee
so that the matrix $t$ is not necessarily Hermitian.
On the state vector space $V$, we can introduce
the basis transformation $U$ by
\be
|h_{j}^{\prime}>=\sum_k |h_k>u_{kj}.\label{btu}
\ee
Under the transformation in Eq. (\ref{btu}), the representation matrix $t$
and the metric matrix $\eta$ transform as follows:
\be
t\longrightarrow t^{\prime}=u^{-1}tu, \;\;\;
\eta\longrightarrow \eta^{\prime}=u^{\dagger}\eta u.\label{bt}
\ee
According to the above statement, we now introduce the important
representation matrices $q$ and $n$ of the Hermitian BRST operator $Q$ and
the anti-Hermitian ghost number operator $N_{gh}$, respectively.
The Hermiticity conditions for $q$ and $n$ assume the following forms:
\bea
& & \eta q=q^{\dagger}\eta,\label{herq}\\
& & \eta n=-n^{\dagger}\eta.\label{hern}
\eea

We shall choose a representation in which $n$ is diagonalized with integral
elements. Note that it is always possible to find an
appropriate $u$ in Eq. (\ref{btu})
which brings $\eta$ into the standard form \cite{Nish84,Bogo90}:
\be
\eta^2=1.\label{stm}
\ee
Equation (\ref{hern}) also shows that the metric matrix $\eta$
defines an isomorphism between the subspaces $V_p$ and $V_{-p}$ of
respective ghost numbers $+p$ and $-p$.
In the BRST cohomology, a natural way to use the Hodge theory argument
is to introduce a new positive-definite
inner product \cite{Witt88} defined by
\be
(x|y)\equiv <x|\eta y> \;\;\; \mbox{for} \;\;|x>,|y> \in V.\label{nip1}
\ee
Note that the new inner product, Eq. (\ref{nip1}), is non-degenerate
by Eq. (\ref{det}) and that the norms of a vector $|x>$ in the state
space $V$ with respect to the new inner product $(\;|\;)$ and the physical
inner product $<|>$ can be expressed as
\bea
& & (x|x)=\sum |x_i|^2,\label{nip2}\\
& & <x|x>=\sum \eta_{ij}x_i^{\ast}x_j.\label{pip}
\eea
Thus, the adjoint operator $\tilde{T}^{\dagger}$ of $T$ in the new metric
defined by $(x|Ty)=(\tilde{T}^{\dagger}x|y)$ satisfies
$\tilde{T}^{\dagger}\equiv \eta T^{\dagger} \eta$, where
$<x|Ty>=<T^{\dagger}x|y>$. Note that the matrix representation
${\tilde t}^{\dagger}$ of $\tilde{T}^{\dagger}$ is equal to the matrix
$t^{\dagger}$. We will replace the notation
$\tilde{T}^{\dagger}$ by $T^{\dagger}$, since we shall treat only the
(anti-)Hermitian operators, so that it raises no confusion.

It will be more obvious later that the metric
matrix $\eta$ has the same roles and philosophy as the Euclidean complex
conjugation ${\bf C}$ in Ref. 5 and the time reversal operation
$T$ in Ref. 8. Since the BRST operator $Q$ cannot be self-adjoint
with respect to the inner product in Eq. (\ref{nip1}),
it is convenient to introduce the
adjoint operator $Q^{\dagger}$ of $Q$ called  the ``co-BRST'' operator
defined by
\be
(Qx|y)=(x|Q^{\dagger}y);\label{defQ}
\ee
its matrix representation is given by Eq. (\ref{herq}):
\be
q^{\dagger}=\eta q \eta.\label{defq}
\ee
Let us introduce the ``Laplacian'' operator $\Delta$ defined by
\be
\Delta=\{Q,Q^{\dagger}\}. \label{laplacian}
\ee
Then one finds that $Q$, $Q^{\dagger}$, and $\Delta$ satisfy the
supersymmetrylike algebra
\be
\{Q,Q^{\dagger}\}=\Delta,\;\;[\Delta, Q]=0,
\;\;[\Delta, Q^{\dagger}]=0.\label{susyal}
\ee

Now the BRST cohomology algebra is given by
\bea
\ba{lll}
&[n,q]=q, &[n,q^{\dagger}]=-q^{\dagger},\\
&q^2=q^{\dagger 2}=0, &\{q,q^{\dagger}\}=\delta,\label{bca}
\ea
\eea
where $\delta$ is a matrix representation of the operator $\Delta$ and
the operators shall be represented
in the basis in which $n$ is diagonalized with integer elements.
Now, we introduce two subspaces of $V$ by
\bea
& & Im\;Q \equiv QV=\{|g>\equiv q|x>\; | \;\; |x> \in V \},\label{imq1}\\
& & Im\;Q^{\dagger} \equiv Q^{\dagger}V=
\{|f>\equiv q^{\dagger}|x>\; | \;\; |x> \in V \}.\label{imq2}
\eea
Due to the nilpotency of $q$ and $q^{\dagger}$, all the states in the BRST
doublet space have zero norms with respect to the physical metric, and
the following properties are satisfied:
\bea
& & q|g>=0,\;\;\;\mbox{for}\;\;|g> \in Im\; Q,\label{nilq1}\\
& & q^{\dagger}|f>=0,\;\;\mbox{for}\;\;|f> \in Im\; Q^{\dagger}.\label{nilq2}
\eea
{}From the positive-definite inner product in Eq. (\ref{nip1}),
\be
(g|g)=\sum|g_i|^2 \neq 0\;\; \mbox{and}\;\;
(f|f)=\sum|f_i|^2\neq 0,\label{norm}
\ee
and Eq. (\ref{norm}) implies that $Q(\eta|g>)=q\eta|g> \neq 0$ and
$Q^{\dagger}(\eta|f>)=q^{\dagger}\eta|f> \neq 0$. These mean that
\be
\eta\; Im\;Q=Im\;Q^{\dagger}\;\;\ \mbox{and}\;\;
\eta\; Im\;Q^{\dagger}=Im\;Q \label{mcp1}
\ee
since the metric matrix $\eta$ is non-singular. Note that the above metric
conjugate pairs must have opposite ghost numbers because of Eq. (\ref{hern}).
On the other hand, $(f|g)=(g|f)=0$.

Next, we define the singlet space
generated by the ``harmonic'' state defined by $Ker\; \Delta$. This singlet
space $V_S$ is equivalent to the statement
\be
V_S=\{|x>\;|\;\; q|x>=q^{\dagger}|x>=0, \;\;|x> \in V \} \label{sins}
\ee
due to the positive definiteness of the new inner product, Eq. (\ref{nip1}).
Then, $V_S$ is orthogonal to $Im\;Q$ and $Im\;Q^{\dagger}$ and satisfies
\be
\eta\;V_S=V_S.\label{mcp2}
\ee
Of course, the metric conjugate pairs in the singlet states also have
opposite ghost numbers.
Therefore, the new positive-definite inner product in Eq. (\ref{nip1})
provides us the orthogonal decomposition about the state space $V$, which
is a sum of linearly independent subspaces
as follows \cite{Nish84,Spie87,Kala91}:
\be
V=V_S \oplus V_D=Ker\;\Delta \oplus Im\;Q \oplus Im\;Q^{\dagger}.\label{hodge}
\ee
This decomposition on the vector space $V$ is just the Hodge
decomposition theorem. According to Eq. (\ref{hodge}),
the BRST doublet space
$V_D$ satisfies the following properties:
\be
Im\;Q=q\;Im\;Q^{\dagger} \;\;\; \mbox{and} \;\;\;
Im\;Q^{\dagger}=q^{\dagger}\;Im\;Q.\label{doub}
\ee
According to Eqs. (\ref{hodge}) and (\ref{doub}),
the condition of the physical subspace
$V_{phys}$ defined by $Ker\;Q$ \cite{Naka90} is equivalent to
\be
V_{phys}=Ker\;\Delta \oplus Im\;Q.\label{phys}
\ee
Thus, the BRST doublet pairs are split in the physical subspace $V_{phys}$.
If the subspace $Ker\;\Delta$ has a positive norm, the physical
space $V_{phys}$ cannot contain a negative norm state because of the
divorce of metric conjugate pairs, Eq. (\ref{mcp1}), so that a state in
$V_{phys}$ is a positive norm or a zero norm.

We define the $p$-th BRST cohomology group in subspace $V_p$ by
the BRST equivalence class of ghost number $p$, that is, the kernel of $Q$
modulo its image:
\bea
H^p(V)  &&\equiv Ker^p Q/Im^p Q, \nonumber\\
        &&\cong Ker^p \Delta, \;\;\;\;\;\;\;\;\;\;
                p=0,\cdots,\pm \mbox{dim}G,\label{coho}
\eea
where $G$ is the structure group of the gauge symmetry under consideration.
The Hodge decomposition theorem directly leads to the isomorphism between
the $p$-th BRST cohomology space $H^p(V)$ and the harmonic state space
$Ker^p\Delta$. Since the metric matrix $\eta$ defines the isomorphism between
the subspaces $V_{p}$ and $V_{-p}$, the metric $\eta$ also induces an
isomorphism in BRST cohomology, namely,
\be
H^p(V) \cong H^{-p}(V).\label{dual}
\ee

With these preliminaries we will study the irreducible representation of
BRST cohomology algebra and the condition of the positivity of the
physical state space in BRST quantization.\\
1. {\it BRST Doublet Representation}

The basis of the BRST doublet representation consists of metric
conjugate pairs in the BRST doublet space, Eq. (\ref{doub}), labeled by the
ghost number:
\bea
& &(N+1, Im\;Q)\equiv(1,0,0,0),\;\;
(N, Im\;Q^{\dagger})\equiv(0,1,0,0),\nonumber\\
& &(-N, Im\;Q)\equiv(0,0,1,0),\;\;
(-N-1, Im\;Q^{\dagger})\equiv(0,0,0,1).\label{drb1}
\eea
The irreducible representation of the BRST cohomology algebra
Eq. (\ref{bca}) is given by
\bea
 \ba{ccc}
  \eta=\left( \ba{cccc}
      0 & 0 & 0 & 1 \\
      0 & 0 & 1 & 0 \\
      0 & 1 & 0 & 0 \\
      1 & 0 & 0 & 0
      \ea \right), \;\;\;
 & n=\left( \ba{cccc}
       N+1 &  0  &  0  &  0 \\
        0  &  N  &  0  &  0 \\
        0  &  0  & -N  &  0 \\
        0  &  0  &  0  & -N-1
       \ea     \right),
 &     \\
 q=a\left( \ba{cccc}
        0 & 1 & 0 & 0 \\
        0 & 0 & 0 & 0 \\
        0 & 0 & 0 & 1 \\
        0 & 0 & 0 & 0
       \ea     \right),  \;\;\;
 & q^{\dagger}=a\left( \ba{cccc}
        0 & 0 & 0 & 0 \\
        1 & 0 & 0 & 0 \\
        0 & 0 & 0 & 0 \\
        0 & 0 & 1 & 0
       \ea     \right), \;\;\;
 & \delta=a^2\left( \ba{cccc}
       1 & 0 & 0 & 0 \\
       0 & 1 & 0 & 0 \\
       0 & 0 & 1 & 0 \\
       0 & 0 & 0 & 1
            \ea     \right)
\ea \label{dmr1}
\eea
where $a$ is a coefficient to be determined.
The matrix representation $\delta$ of the
Laplacian operator $\Delta$ in Eq. (\ref{dmr1}) shows us
that no quartet member appears in the BRST cohomology
space: {\it Quartets are always confined!}\\
2. {\it BRST Singlet Representation}\\

There are two kinds of singlet representations in the space,
Eq. (\ref{sins}), with the ghost numbers $N=0$ and $N\neq0$.
For convenience, we consider the two cases together:
\be
(0, Ker\;\Delta),\;\;(N, Ker\;\Delta),\;\;(-N, Ker\;\Delta).\label{srb}
\ee
The irreducible representation of the BRST
cohomology algebra, Eq. (\ref{bca}), is given by
\bea
 \ba{ccc}
  \eta=\left( \ba{ccc}
      1 & 0 & 0  \\
      0 & 0 & 1  \\
      0 & 1 & 0
       \ea \right), \;\;\;
 & n=\left( \ba{ccc}
        0  &  0  &  0   \\
        0  &  N  &  0   \\
        0  &  0  & -N
        \ea     \right),
&    \\
q=\left( \ba{ccc}
       0 & 0 & 0  \\
       0 & 0 & 0  \\
       0 & 0 & 0
       \ea     \right),  \;\;\;
 & q^{\dagger}=\left( \ba{ccc}
       0 & 0 & 0 \\
       0 & 0 & 0 \\
       0 & 0 & 0
       \ea     \right), \;\;\;
 & \delta=\left( \ba{ccc}
       0 & 0 & 0 \\
       0 & 0 & 0 \\
       0 & 0 & 0
         \ea     \right).
   \ea \label{smr}
\eea

{}From the Eqs. (\ref{dmr1}) and (\ref{smr}), we see that
the dimension of the harmonic state space $Ker\;\Delta$ is not less
than $Tr\;\eta$  and that the dimension
of the space $V_S^{(+)}$ generated by the basis $(0,Ker\;\Delta)$ is equal
to the index $Tr\;\eta$. If $Q$ is complete, that is, the set of all the
zero-norm states in $Ker\;Q$ is $Im\;Q$ \cite{Spie87},
the completeness condition of $Q$ is obviously equivalent to the condition
of the absence of the paired singlet with $N\neq0$.
If this condition is satisfied, the norms of all the states in $Ker\;Q$ must
have the same sign, as shown in Ref. 5. Therefore, we can
always choose the sign of $Ker\;Q$ as positive. Then $Tr\;\eta$ is equal to
the dimension of the positive-definite Hilbert space as claimed in Ref. 5.
In addition, the ``split a pair'' mechanism is complete in
$V_{phys}$ so that negative norm states cannot be generated in $V_{phys}$.
Consequently, the BRST cohomology space is 0-norm-state free, thus proving
the ``no-ghost'' theorem.
We, thus, conclude that the cohomology space $H(V)$ isomorphic with the
harmonic space $Ker\;\Delta$ has a positive-definite metric so that the
physical Hilbert space has a positive-definite norm as long as the paired
singlet is absent.

In order to show how the illustrative example in Ref. 5 can be
realized through the quartet mechanism
in the BRST doublet representation, Eq. (\ref{dmr1}),
it will be interesting to take the basis
transformation, Eq. (\ref{btu}), into the basis
for the metric matrix $\eta$ to be diagonalized. Then, the basis of the
BRST doublet representation consists of the metric eigenstates
\bea
& & \vec{k}\equiv\frac{1}{2}(1,-1,1,-1),\;\;
\vec{l^{\prime}}\equiv\frac{1}{2}(-1,-1,1,1),\nonumber\\
& & \vec{l}\equiv\frac{1}{2}(1,1,1,1),
\;\;\vec{k^{\prime}}\equiv\frac{1}{2}(-1,1,1,-1).\label{drb2}
\eea
The irreducible representation of the BRST cohomology algebra is given by
\bea
&& \eta=\left( \ba{cccc}
      -1 & 0 & 0 & 0 \\
       0 &-1 & 0 & 0 \\
       0 & 0 & 1 & 0 \\
       0 & 0 & 0 & 1
      \ea \right), \;\;
 n=\left (\ba{cccc}
        0          &     0         & \frac{1}{2}  &-N-\frac{1}{2}  \\
        0          &     0         & -N-\frac{1}{2} & \frac{1}{2} \\
    \frac{1}{2}    & -N-\frac{1}{2}&      0         &      0 \\
    -N-\frac{1}{2} & \frac{1}{2}   &      0         &      0
           \ea     \right), \label{dmr2}
  \\
 &&q=\frac{a}{2} \left ( \ba{cccc}
                    -1 & 0 & 1 &  0 \\
                     0 & 1 & 0 & -1 \\
                    -1 & 0 & 1 &  0 \\
                     0 & 1 & 0 & -1
                       \ea     \right),  \;\;
  q^{\dagger}=\frac{a}{2}\left( \ba{cccc}
       -1 &  0 & -1 &  0 \\
        0 &  1 &  0 &  1 \\
        1 &  0 &  1 &  0 \\
        0 & -1 &  0 & -1
          \ea     \right), \;\;
  \delta=a^2\left( \ba{cccc}
       1 & 0 & 0 & 0 \\
       0 & 1 & 0 & 0 \\
       0 & 0 & 1 & 0 \\
       0 & 0 & 0 & 1
            \ea     \right).
 \nonumber
\eea
Then, analogies with the example in Ref. 5 can be
definitely realized through the quartets in Eq. (\ref{dmr2}) in the BRST
cohomology. Of course, the same words
can also be applied to the representation in Eq. (\ref{dmr1}).
First, notice the following facts:
\bea
& & n\vec{k}=(N+1)\vec{k},\;\;n\vec{k^{\prime}}=
-(N+1)\vec{k^{\prime}},\nonumber\\
& & n\vec{l}=-N\vec{l},\;\;n\vec{l^{\prime}}=N\vec{l^{\prime}},\label{ghon}\\
& & \eta\vec{k}=\vec{k^{\prime}},\;\;\;
\eta\vec{l}=\vec{l^{\prime}}.\label{mkl}
\eea
Second, the matrices $q$ and $q^{\dagger}$ can be written as
\be
q=a(\tensor {kl}+\tensor {lk}),
\;\;q^{\dagger}=\eta q \eta=
a(\tensor {k^{\prime} l^{\prime}}
+\tensor {l^{\prime} k^{\prime}}) \label{q}
\ee
where $\tensor {kl}\equiv k^i l_j$. Then, we easily find that
\bea
& & Ker\;Q=\{\vec{k}=\frac{1}{a}q\vec{l^{\prime}},\;\;
\vec{l}=\frac{1}{a}q\vec{k^{\prime}}\}=Im\;Q,\nonumber\\
& & Ker\;Q^{\dagger}=\{\vec{k^{\prime}}=\frac{1}{a}q^{\dagger}\vec{l},\;\;
\vec{l^{\prime}}=\frac{1}{a}q^{\dagger}\vec{k}\}=Im\;Q^{\dagger}.\label{kerq}
\eea
Therefore, if the singlet pairs with non-zero ghost number are absent,
the completeness of $Q$ and the ``split a pair'' mechanism are also obvious
in the representation of Eq. (\ref{dmr2}). Slight differences with the
Ref. 5 exist in the above identifications about quartet states
in that our members are more appropriate since $q$ increases the ghost number
of a state by one unit while $\eta$ connects states with opposite
ghost numbers. In the next section, we will study the explicit
example of the BRST cohomology discussed in Sec. II in the context of QED.

\section{BRST COHOMOLOGY IN QED}
\label{sec:QED}

Consider the (anti-)BRST invariant effective QED Lagrangian.
\bea
{\cal L}_{eff} = &-&\frac{1}{4}F_{\mu\nu}F^{\mu\nu}+ {\bar\psi}
(i\gamma^{\mu}D_{\mu}-m)\psi-\frac{1}{2}
{\bar s}s(A_{\mu}^2+\alpha{\bar c}c) \nonumber\\
= &-&\frac{1}{4}F_{\mu\nu}F^{\mu\nu}+ {\bar\psi}
(i\gamma^{\mu}D_{\mu}-m)\psi + A_{\mu}\partial^{\mu}b+\frac{\alpha}{2}b^2
-\partial_{\mu}{\bar c}\partial^{\mu}c \label{qedl}
\eea
where $D_{\mu}= \partial_{\mu}+ieA_{\mu}$ is the covarint derivative with
the metric $g_{\mu\nu}=(1,-1,-1,-1).$ (Our BRST treatments are parallel
with those of Baulieu's paper \cite{Baul85}.)
This effective Lagrangian has the rigid symmetry under the following BRST
transformation:
\bea
 & sA_{\mu}=\partial_{\mu}c,\;\;\;  & sc=0,\nonumber\\
 & s{\bar c}=b,\;\; &  sb=0,\label{qedt}\\
 & s\psi=-iec\psi.\nonumber
\eea
We introduced an auxiliary field $b$  to achieve off-shell nilpotency of the
BRST transformation.
Then, the nilpotent conserved N$\ddot{o}$ther charges generated by
the BRST transformation in Eq. (\ref{qedt}) read as
\be
Q = \int d^3x \{-(\partial_i F^{io}-\rho)c-b\dot{c}\} \label{qedq}
\ee
where $\rho$ is a charge density defind by
\be
\rho=e {\bar \psi}\gamma_0 \psi.\label{qedch}
\ee
The transformations in Eq. (\ref{qedt}) now can be defined as follows:
$s{\cal F}(x)=i[Q,{\cal F}(x)\}$ where the symbol $[\;,\;\}$ is the
graded commutator.
In the language of quantum field theory, the BRST operator $Q$ is the
generator of the quantum gauge transformation.

For simplicity, we only consider
the free Maxwell theory since the matter fields are not essential in the
BRST cohomology. Using the mode expansion of field
operators \cite{Henn92},
we find the following expression about the BRST operator $Q$:
\be
Q=\sum_{{\bf k}} (a_{{\bf k}}c_{{\bf k}}^{\dagger}
+a_{{\bf k}}^{\dagger}c_{{\bf k}}) \label{modq1}
\ee
where $a_{{\bf k}}(a_{{\bf k}}^{\dagger})$ is a linear combination
of the longitudinal photon $a_3(a_3^{\dagger})$ and the temporal
photon $a_0(a_0^{\dagger})$ of momentum ${\bf k}$ and is given by
$a_{{\bf k}}=a_{3{\bf k}}-a_{0{\bf k}}\; (a_{{\bf k}}^{\dagger}=
a_{3{\bf k}}^{\dagger}-a_{0{\bf k}}^{\dagger})$ with its conjugate variable
$b_{{\bf k}}^{\dagger}\equiv\frac{1}{2}(a_{3{\bf k}}^{\dagger}+
a_{0{\bf k}}^{\dagger})\; (b_{{\bf k}}\equiv\frac{1}{2}(a_{3{\bf k}}+
a_{0{\bf k}}))$.
Notice canonical quantization leads to
\bea
[a_{{\bf k}},b_{{\bf k}^{\prime}}^{\dagger}]=
\delta_{{\bf k}{{\bf k}^{\prime}}},\;\;
[b_{{\bf k}},a_{{\bf k}^{\prime}}^{\dagger}]=
  \delta_{{\bf k}{{\bf k}^{\prime}}},\nonumber\\
\{c_{{\bf k}},{\bar c}_{{\bf k}^{\prime}}^{\dagger}\}=
  \delta_{{\bf k}{{\bf k}^{\prime}}},\;\;
\{{\bar c}_{{\bf k}},c_{{\bf k}^{\prime}}^{\dagger}\}=
  \delta_{{\bf k}{{\bf k}^{\prime}}};\label{comm}
\eea
the other (anti-)commutators vanish.

Now, we will apply the BRST cohomology described in Sec. II to QED. We define
the physical vacuum $|0>$ as the state annihilated by all the destruction
operators:
\be
a_{{\bf k}}|0>=b_{{\bf k}}|0>=c_{{\bf k}}|0>=
{\bar c}_{{\bf k}}|0>=0.\label{vacuum}
\ee
We construct the Fock space $\Omega$ which has an indefinite metric and is
decomposed into the tensor product
\[ \Omega=\Omega_T \otimes \Omega_F.\]
Here, $\Omega_T$ is the space $V_S^{(+)}$ discussed in Sec. II where,
for example, a transverse photon lives, and the subspace $\Omega_F$ of
the unphysical sectors is generated by
\be
|m,n;k,l>\equiv\frac{1}{\sqrt{m!n!}}
a^{\dagger m}b^{\dagger n}c^{\dagger k}{\bar c}^{\dagger l}|0>.\label{fockst}
\ee
Since the conjugate pairs of the (anti-)commutators in Eq. (\ref{comm})
will be adjoints of one another under the inner product in Eq. (\ref{nip1}),
one can easily find the following relations:
\bea
&& \eta a_{{\bf k}} \eta =b_{{\bf k}},\;\;\;
\eta a_{{\bf k}}^{\dagger} \eta= b_{{\bf k}}^{\dagger},\nonumber\\
&& \eta c_{{\bf k}} \eta={\bar c}_{{\bf k}},\;\;\;
\eta c_{{\bf k}}^{\dagger} \eta={\bar c}_{{\bf k}}^{\dagger}. \label{etacon}
\eea
Then, the explicit form of the adjoint operator $Q^{\dagger}$ of $Q$ is
\be
Q^{\dagger}=\sum_{{\bf k}} (b_{{\bf k}}{\bar c}_{{\bf k}}^{\dagger}
+b_{{\bf k}}^{\dagger}{\bar c}_{{\bf k}}).\label{modq2}
\ee
This operator $Q^{\dagger}$ is consistent with the definition
in Eq. (\ref{defQ}) for the ``co-BRST'' operator.
The ``Laplacian'' operator $\Delta$ defined in Sec. II can then be
described by
\be
\Delta=\sum_{{\bf k}} (a_{{\bf k}}^{\dagger}b_{{\bf k}}+
b_{{\bf k}}^{\dagger}a_{{\bf k}}+{\bar c}_{{\bf k}}^{\dagger}c_{{\bf k}}+
c_{{\bf k}}^{\dagger}{\bar c}_{{\bf k}}),\label{moddel}
\ee
and this is the number operator $N_{unphys}$ for the unphysical modes.
Thus, the operator $\Delta$ is essentially the part of the Hamiltonian with
the unphysical fields if it is multiflied by the
frequency $\omega_{{\bf k}}$ for each mode of momentum ${\bf k}$.
This property of our BRST cohomology is
reminiscent of the definition in Ref. 8 about a quantum cohomology
whose cohomology classes correspond to quantum ground states.
{}From Eq. (\ref{moddel}), we can find the coefficient $a$ in
Eq. (\ref{dmr1}) or Eq. (\ref{dmr2}); this is just the number of
unphysical particles. The quartet members are characterized by the
eigenvalues of the
number operator $N_{unphys}$.

Let us introduce the ``total'' ghost number operator $N_{gh}^{total}$
defined by
\bea
&&N_{gh}^{total}=N_{gh}^{(0)}+N_{gh}^{(1)},\nonumber\\
&&N_{gh}^{(0)}=\sum_{{\bf k}} (a_{{\bf k}}^{\dagger}b_{{\bf k}}-
b_{{\bf k}}^{\dagger}a_{{\bf k}}),\;\;\;
N_{gh}^{(1)}=\sum_{{\bf k}}({\bar c}_{{\bf k}}^{\dagger}c_{{\bf k}}-
c_{{\bf k}}^{\dagger}{\bar c}_{{\bf k}}) \label{totgh}
\eea
where $N_{gh}^{(1)}$ is just the ordinary ghost number operator $N_{gh}$
introduced in Sec. II. The states $|m,n;k,l>$ in Eq. (\ref{fockst}) are
eigenstates of $N_{gh}^{(0)}$, $N_{gh}^{(1)}$ and $\Delta$ with eigenvalues
of $m-n$, $k-l$, and $m+n+k+l$, respectively.
{}From Eq. (\ref{totgh}) and Eq. (\ref{comm}), one can easily find the
following properties:
\bea
&&[N_{gh}^{(0)}, Q]=Q,\;\;\;[N_{gh}^{(1)}, Q]=Q,\nonumber\\
&&[N_{gh}^{(0)}, Q^{\dagger}]=-Q^{\dagger},\;\;\;
  [N_{gh}^{(1)}, Q^{\dagger}]=-Q^{\dagger},\nonumber\\
&&[N_{gh}^{total}, Q]=2Q,\;\;\;
  [N_{gh}^{total}, Q^{\dagger}]=-2Q^{\dagger},\;\;\;
  [N_{gh}^{total}, \Delta]=0. \label{ghcomm}
\eea

The bases corresponding to the BRST doublet representation in Eq.(\ref{drb1})
consist of the following Fock states (using the same
notations defined by Eqs. (\ref{drb1}) and (\ref{drb2})):
\bea
(I):\;\;&& |0,m;1,0>=(1,0,0,0),\;|0,m+1;0,0>=(0,1,0,0),\nonumber\\
&&|m+1,0;0,0>=(0,0,1,0),\;|m,0;0,1>=(0,0,0,1), \label{fbas1}\\
(II):\;\;&& |n+1,m;1,0>=(1,0,0,0),\nonumber\\
&&\sqrt{\frac{m+1}{n+m+2}}|n+1,m+1;0,0>
-\sqrt{\frac{n+1}{n+m+2}}|n,m;1,1>=(0,1,0,0),\nonumber\\
&&\sqrt{\frac{m+1}{n+m+2}}|m+1,n+1;0,0>
+\sqrt{\frac{n+1}{n+m+2}}|m,n;1,1>=(0,0,1,0),\nonumber\\
&&|m,n+1;0,1>=(0,0,0,1).\label{fbas2}
\eea
The bases (I) in Eq. (\ref{fbas1}) and (II) in Eq. (\ref{fbas2}),
respectively, correspond to the eigenvalues $a^2=m+1$ and $a^2=n+m+2$ of
$N_{unphys}$ or $\Delta$ in the representation of Eq. (\ref{dmr1}).
They exhaust all zero-norm states in $Ker Q$ and form the quartets in
the (m+1)- and (n+m+2)-unphysical sectors:
\begin{eqnarray}
 \ba{ccc}
  \eta=\left( \ba{cccc}
      0 & 0 & 0 & 1 \\
      0 & 0 & 1 & 0 \\
      0 & 1 & 0 & 0 \\
      1 & 0 & 0 & 0
      \ea \right), \;\;\;
   n=\left( \ba{cccc}
        1  &  0  &  0  &  0 \\
        0  &  0  &  0  &  0 \\
        0  &  0  &  0  &  0 \\
        0  &  0  &  0  & -1
       \ea     \right),\;\;\;
 & n^{(0)}_I=\left( \ba{cccc}
        -m &  0   &  0   &  0 \\
        0  & -m-1 &  0   &  0 \\
        0  &  0   & m+1  &  0 \\
        0  &  0   &  0   &  m
       \ea     \right),
 &     \\
 q=\left(\ba{c}
          \sqrt{m+1} \\
	  \sqrt{n+m+2}
	 \ea   \right)
\left( \ba{cccc}
        0 & 1 & 0 & 0 \\
        0 & 0 & 0 & 0 \\
        0 & 0 & 0 & 1 \\
        0 & 0 & 0 & 0
       \ea     \right),  \;\;\;
 & q^{\dagger}=\left(\ba{c}
          \sqrt{m+1} \\
	  \sqrt{n+m+2}
	 \ea   \right)
\left( \ba{cccc}
        0 & 0 & 0 & 0 \\
        1 & 0 & 0 & 0 \\
        0 & 0 & 0 & 0 \\
        0 & 0 & 1 & 0
       \ea     \right),
 &  \\
 \delta=\left(\ba{c}
          m+1 \\
	  n+m+2
	 \ea   \right)
\left( \ba{cccc}
       1 & 0 & 0 & 0 \\
       0 & 1 & 0 & 0 \\
       0 & 0 & 1 & 0 \\
       0 & 0 & 0 & 1
            \ea     \right).
\ea \label{fmr}
\end{eqnarray}
$n^{(0)}_I$ in Eq. (\ref{fmr}) is the matrix representation for states
$(I)$ of the operator $N_{gh}^{(0)}$ in Eq. (\ref{totgh}), and the
representation for states $(II)$ can be obtained by the replacement
$m\longrightarrow m-n-1$.
The matrix structures of the representation in Eq. (\ref{fmr}) clearly
show consistent results, such as the isomorpism between $V_p$ and $V_{-p}$
and the ``split a pair'' and quartet mechanisms discussed in Sec. II.
Therefore, these quartet members disappear in the BRST cohomology.
No paired singlet exists and so the BRST singlet states come only
from the harmonic states in $(0,Ker \Delta)$.
Consequently, the physical Hilbert space $V_S$ is
expressed by the Fock space $\Omega=\Omega_T \otimes |0>_F$
which has a positive-definite norm.
In the metric-diagonal basis of Eq. (\ref{drb2}),
we also ensure these conclusions
along the equivalent logics, Eqs. (\ref{ghon})-(\ref{kerq}).

\section{DISCUSSION}
\label{sec:CON}
We have shown that there is a straightforward way to isolate
the physical Hilbert
space with a positive-definite metric through the BRST cohomology.
Using the co-BRST operator, we have obtained the Hodge decomposition
theorem under a positive-definite inner product and have uniquely chosen the
harmonic representative with a positive-definite norm as long as
$Q_{BRST}$ is complete.

In general, the BRST cohomology algebra cannot prevent the invasion
of a paired singlet with non-zero ghost number, and the ``physical'' meaning
of the higher cohomology groups $H^p(V),\;p\neq0$, is not obvious.
If the nontrival higher cohomologies appeared in the theory, it would be
a very interesting problem to give their ``physical'' interpretations.

Our BRST cohomology is quite different from the cohomology in the recent
literature \cite{Holt90,Lave93}, which cannot be applied to
the problem to directly isolate the physical state with positive-definite
norm. Our definition about a ``co-BRST'' operator is quite similar to
the ``dual BRST'' operator in Ref. 12.

Our ``co-BRST'' operator defined in QED is not Lorentz invariant
because we have flipped the sign of the time-directional operators,
but it is a conserved operator that commutes
with Hamiltonian, so our recipes for characterizing the physical Hilbert
space using the Hodge decomposition theorem have physical significance.
An explicit construction of the BRST cohomology in Fock space
when the (self-)interaction is present is, in general, a very difficult
problem and we have no solutions to this problem. However, see, for example,
Ref. 13 which analyzed the unitarity of the S-matrix and the positivity
of the physical states norm in the subspace
of asymptotic states under the assumption about the asymptotic completeness.
Reference 14 uses a similar structure to that in our approach to
construct the state space in the BRST quantization, but the state space
is not completely characterized because they did not use the some explicit
form of the metric as ours.

We also found \cite{Yang} the local, but
non-covariant, symmetries in Abelian gauge theories.
The BRST-like charge $Q^{\perp}$ constructed
in Ref. 11 corresponds to the ``co-BRST'' operator in
Ref. 10, and there is no reason to
abandon locality, unlike the claim of the Ref. 11.

\section*{ACKNOWLEDGEMENTS}
This work was supported by the Korean Science and Engeneering Foundation
(94-1400-04-01-3 and CTP) and
by the Korean Ministry of Education (BSRI-94-2441).

\end{document}